\documentclass{article}

\usepackage{authblk}

\usepackage{amssymb}

 \usepackage{lineno}

\usepackage{amsmath} 
\usepackage{graphicx}
\usepackage{listings}
\usepackage{mathtools}
\usepackage{longtable}
\usepackage{hyperref}
\usepackage{color}

\newcommand{\bfv}{\mathbf{v}}

\newcommand{\bfx}{\mathbf{x}}

\title{Comparison of Preconditioning Strategies in Energy Conserving Implicit Particle in Cell Methods}

\author[1]{Siddi Lorenzo}
\author[2]{Cazzola Emanuele}
\author[1]{Lapenta Giovanni}

\affil[1]{Department of Mathematics, KULeuven, University of Leuven, Belgium (EU)}
\affil[2]{Department of Mathematics, University of Surrey, Guildford, Surrey, UK}

\date{}                     

\begin{document}

\maketitle

\lstset{language=Matlab,frame=single,
    breaklines=true,%
    morekeywords={matlab2tikz},%
    basicstyle=\ttfamily\scriptsize}
    

\begin{abstract}
This work presents a set of preconditioning strategies able to significantly accelerate the performance of fully implicit  energy-conserving Particle-in-Cell methods to a level that becomes competitive with semi-implicit methods. 
We compare three different preconditioners. We consider three methods and compare them with a straight unpreconditioned Jacobian Free Newton Krylov (JFNK) implementation. 
The first two focus, respectively, on improving the handling of particles (particle hiding) or fields (field hiding) within the JFNK iteration. 
The third uses the field hiding preconditioner within a direct Newton iteration where a Schwarz-decomposed Jacobian is computed analytically. Clearly, field hiding used with JFNK or with the direct Newton-Schwarz (DNS) method outperforms all method.
We compare these implementations with a recent semi-implicit energy conserving scheme. Fully implicit methods are still lag behind in cost per cycle but not by a large margin when proper preconditioning is used. However, for exact energy conservation, preconditioned fully implicit methods are significantly easier to implement compared with semi-implicit methods and can be extended to fully relativistic physics. 
\end{abstract}




\section{Introduction}
Particle in cell plasma simulations \cite{birdsall-langdon,hockney-eastwood} are amongst the most successful in obtaining new physics and in running at top performance on supercomputers \cite{vpic}. Recently, this success has attracted several new developments. Among them, we focus the attention here on implicit and semi-implicit PIC methods.
Given their simplicity and flexibility, these
PIC formulations have gained wide use as effective computational tools for
the simulation of multiscale plasma phenomena in a variety of fields from space \cite{lapenta2012particle}, 
astrophysics \cite{sironi2013}, laser-plasma interaction \cite{silva2004}, 
magnetic fusion \cite{candy2003} and space propulsion \cite{cazzola2016}.
The interest on (semi-)implicit PIC has been intensified by the presentation of exactly energy conserving schemes \cite{markidis2011energy,chen2011energy}. 
Semi-implicit methods had been previously developed along two lines of research (e.g. \cite{Brackbill:1985,lapenta-book:2017} for a review):  \emph{moment implicit} \cite{brackbill-forslund} and \emph{direct implicit} \cite{directimplicit}. In these methods, the coupling between particles and fields is represented via a linear response: the direct implicit method uses a sensitivity matrix while the implicit moment method uses a Taylor expansion similar to the Sonine polynomial expansion of the Chapman-Enskog moment method \cite{chapman1970mathematical}. Both methods found very successful applications either in commercial codes (\url{www.orbitalatk.com/lsp/}) or in research codes used by the plasma physics community: Venus\cite{brackbill-forslund}, Celeste\cite{vu,lapenta05}, Parsek2D \cite{parsek} and iPic3D \cite{ipic3d}. 

Very recently, a new variant of semi-implicit method, ECsim \cite{Lapenta2017349} has been proposed. The new method is based on constructing a so-called mass matrix that represents how the particles in the system produce a plasma current in response to the presence of magnetic and electric fields. This response is linear and the ECsim method belongs to the class of semi-implicit methods. However, its key difference compared with the previously mentioned semi-implicit methods is that its mass matrix formation leads to exact energy conservation. A theorem proves the energy conservation to be valid at any finite time step\cite{Lapenta2017349}, a conclusion confirmed to machine precision in practical multidimensional implementations\cite{lapenta2017multiple}. 

Fully implicit methods had been deemed impractical in the past decades, but recent developments in handling non linear coupled systems made them practical. The original particle mover and field solver of the implicit moment method were already exactly energy conserving\cite{markidis2011energy}. This property is however broken by the Taylor expansion. If instead the discretised field and particle equations are solved directly as a coupled non-linear system, energy is conserved exactly \cite{markidis2011energy}. Several variants can include charge conservation besides energy conservation\cite{chen2011energy}, for electromagnetic relativistic systems \cite{lapenta2011particle} and for the so-called Darwin pre-Maxwell approximation \cite{chen2015multi}.

Even in our times, non-linear systems still are a formidable numerical task, especially for a large number of unknowns. For an efficient implementation, the numerical method used for solving the system needs to be carefully designed. So far, the research has focused on the Jacobian-Free Newton Krylov method (JFNK) \cite{kelley} and on Picard iterations \cite{taitano2013development}. The performance of the non linear solver can be improved dramatically if good preconditioners are designed \cite{knoll2004jacobian}. Preconditioners are operators that invert part of the problem, increasing the speed of convergence of the solution of the Jacobian problem needed for each Newton iteration. 

In the context of implicit PIC, the attention has focused on the so-called particle hiding (or particle enslavement) approach \cite{kim,markidis2011energy}, especially suitable to hybrid architectures \cite{chen2012efficient},  and the use of fluid preconditioning for the fields \cite{chen2014fluid}.

In the present paper we revisit the particle hiding preconditioning and compare it with a number of  new preconditioning methods able to significantly accelerate the iterative computation.

The first new approach proposed has been named \emph{field hiding} after its similarities with the particle hiding method, with the main difference being its acting on the field solver instead of on the particle mover. 

Second, we show that this method can be further optimized in PIC methods if the JFNK is completely replaced by a direct Newton solver where a Schwarz decomposed Jacobian is computed and inverted analytically, a method we refer to as Direct Newton-Schwarz (DNS). 

Finally, all the implementations of the fully implicit method described above are compared with the new semi-implicit method ECsim.  

All methods considered (including ECsim) conserve energy exactly, to machine precision and are virtually identical in the accuracy of the results produced for the cases tested. They only differ in the computing performance. We focus then on comparing the different implementations in terms of CPU time used.
In particular, the analysis is conducted in the form of four tests to measure the computing performance while: 1) varying the time step, 2) varying the  tolerance of the NK iterations, 3) varying the spatial grid resolution and 4) varying the number of particles.

We structured the paper as follows. Section \ref{sec:methods} gives an introduction to the formulation of the energy conserving fully implicit method. Section \ref{sec:impl}  describes the preconditioning methods mentioned earlier together with a skeleton of their  implementation in MATLAB. The results of the analysis are given in Section \ref{sec:results}. Finally,
Section \ref{sec:conclusions} provides the discussion of the results and some important conclusion.

\section{Base Method for Comparative Testing:  Energy Conserving Fully Implicit Method (ECFIM)} \label{sec:methods}

For 1D electrostatic problems, the ECFIM has been previously presented in terms of the Vlasov-Amp\`ere system \cite{markidis2011energy, chen2011energy}, but it can be equivalently expressed in terms of the Vlasov-Poisson system \cite{chen2011energy}, which is
the approach followed in this work. We illustrate here the application in 1D but for all methods presented there is no barrier to the extension to multiple dimension, other than coding complexity.

The particle-in-cell method is based on introducing a new entity called computational particle, or \emph{super-particle}, which is intended to represent a particular cluster of physical particles ($w_p$), having a charge $q_{p}=w_pq_{s}$, where $q_{s}$ is the charge of the physical particles of species $s$. Similarly, the mass is $m_{p}=w_pm_{s}$. The charge to mass ratio is physically correct and the (super)-particles still move with the same Newton equations of the physical particles. 
The core idea of the ECFIM is to use an implicit time centered  particle mover: 
\begin{equation} \label{eq:PICeq}
\begin{split}
  &{x}_p^{n+1} = {x}_p^n + \bar{{v}}_p \Delta t \\
  &\bar{{v}}_p = {v}_p^n + \frac{q_s}{2 m_s} \Delta t \bar{{E}}_p.   
  \end{split}
\end{equation}
where quantities under bar are averaged between the time step $n$ and $n+1$ (e.g. $\bar{{v}}_p = ({v}_{p}^{n}  +{v}_{p}^{n+1})/2$). The new velocity at the advanced time is then simply:
\begin{equation} \label{eq:ECPICmover_av}
{v}_p^{n+1} = 2 \bar{{v}}_p - {v}_p^n.
\end{equation}   
The electric field  $\bar{E}_p$ is the  electric field acting of the computational particle $p$ with position ${x}_p$ and velocity ${v}_p$ at the average time $\bar{{E}}_p = ({E}_{p}^{n}  +{E}_{p}^{n+1})/2$.
Unlike the Vlasov-Amp\`ere method, we  rely here on the Poisson's equation to compute the electric field:
 \begin{equation} \label{eq:maxwell}
\epsilon_{0} \frac{\partial^2 \phi}{\partial x^2} =  - \rho \end{equation}
 where $\rho$ is the charge density. However, to reach an exactly energy conserving scheme, we reformulate the equation using explicitly the equation of charge conservation:
  \begin{equation} \label{eq:chcons}
\frac{\partial \rho}{\partial t} = - \frac{\partial J}{\partial x}.
\end{equation}
We take the temporal partial derivative of the Poisson's equation (\ref{eq:maxwell}) and substitute eq. (\ref{eq:chcons})
\begin{equation} \label{eq:field}
\epsilon_{0} \frac{\partial^2 }{\partial x^2}\frac{\partial \phi}{\partial t} =   \frac{\partial J}{\partial x}.
\end{equation}
Assuming now a time centered discretization, we express the variation of the scalar potential in one time step, $\delta \phi = \frac{\partial \phi}{\partial t}\Delta t$ as
\begin{equation} \label{eq:field-discr}
\epsilon_{0} \frac{\partial^2 \delta \phi }{\partial x^2} =   \frac{\partial J}{\partial x} \Delta t
\end{equation}
and compute the new electric field as: 
\begin{equation} \label{eq:newE}
E^{n+1} = E^n -\nabla \delta \phi
\end{equation}
This equation can then be further discretized in space. We use here a staggered 1D discretization where the electric field and current are computed in the nodes of a uniform grid and the potential in the centers. In 1D this leads to $x_c=\{(i+1/2)*\Delta x\}$ for $i \in [0,N_g]$
 and $x_v=\{i*\Delta x\}$ for $i \in [0,N_g+1]$ where $N_g$ is the number of cells. The fully discretized field equation is then:
 \begin{equation} \label{eq:field-discr-discr}
\epsilon_{0} (\delta \phi_{c+1} +\delta \phi_{c-1} -2 \delta \phi_{c}) =     (J_{v+1}-J_v) \Delta t \Delta x
\end{equation}
where $v$ is the node to the left of the center $c$. The time dependent Poisson formulation is Galilean invariant and does not suffer from the presence of any curl component in the current \cite{chen2011energy}. In electrostatic systems the current cannot develop a curl because such curl would develop a corresponding curl of the electric field, and in consequence electromagnetic effects. The formulation above prevents that occurrence since it is based on the divergence of the current and any  curl component of J is eliminated. This is not an issue in 1D but it would be in higher dimensions: the Vlasov-Poisson system can be more directly applied for electrostatic problems. 

The peculiarity of PIC methods is the use of  interpolation functions $W \left( {x}_g - \bar{x}_p \right)$ ($g$ the generic grid point, center or vertex) to describe the coupling between particles and fields. 
For example,  in the case of a  the Cloud-in-Cell approach, the interpolation function reads~ \cite{hockney-eastwood}
\begin{equation} \label{eq:interp}
 W\left( {x}_g - \bar{x}_p \right)=\begin{cases}
        1 - \frac{| {x}_g - \bar{x}_p |}{\Delta x}, & \text{if } | {x}_g - \bar{x}_p | < \Delta x \\
        0, & \text{otherwise. } \\
        \end{cases}
\end{equation}
The interpolation functions can then be used to compute the electric field acting on each computational particles as
\begin{equation}
  {E}_p = \sum_v = {E}_g W\left( {x}_g - \bar{x}_p \right)  \label{eq:defE}
  \end{equation}
where ${E}_g$ is the electric at the vertices of the grid.
  Likewise, the current is computed as
\begin{equation} \label{eq:current} 
 \bar{{J}}_v = \frac{\sum_p q_p \bar{{v}}_p W\left( {x}_g - \bar{x}_p \right)}{\Delta x}\end{equation}
The set of equations for particles (\ref{eq:PICeq}) and fields (\ref{eq:field-discr-discr})  is non-linear and coupled. The coupling comes from the need to know the electric field at the advanced time before the particles can be advanced and in the need to know the new particles position and velocity before the current can be computed and the field advanced. The set of equations is also non-linear due to the non-linear dependence of the currents and fields on the particle positions. A non-linear solver is then needed to address the problem.

The solution provided describes an exactly energy conserving scheme \cite{markidis2011energy, chen2011energy}: energy will be conserved for any choice of the time step $\Delta t$ at the level of accuracy chosen in the interactive solution of the system equations.

\section{Skeleton MATLAB Implementations of the Preconditioners for the ECFIM} \label{sec:impl}
The goal of the study is to compare under identical straightforward style of coding  different strategies for conducting the non-linear iteration required by the fully implicit method. To this end, we rely on a MATLAB implementation that we include in the appendices.  Naturally, this is a limited scope: more sophisticated implementations in other languages for more complex problems would likely be needed for real world applications. Our aim here is to consider the promise shown in the simplest cases by each of the following strategies for the ECFIM:
\begin{enumerate}
\item  Jacobian Free Newton Krylov (JFNK) method without any preconditioning. 
\item JFNK with particle hiding, also called particle enslavement in the literature  \cite{markidis2011energy,chen2011energy}.
\item JFNK with field hiding 
\item Direct Newton approach with field hiding and using a partial Jacobian  computed in a Schwarz particle-by-particle decomposition. We refer to this new method as Direct Newton-Schwartz (DNS).
\end{enumerate}
These implementations are additionally compared  with the ECSIM method \cite{Lapenta2017349,lapenta2017multiple}, which  does not require any non linear iteration and is  used here as comparative benchmark.

In conclusion, we would like to point out that
 the results obtained here in a simple case might be overturned in more complex cases using more advanced implementations. Nevertheless, we believe this study may be considered as a hitchhiker guide for more complex design choices in production codes.
Below we report the details of the implementation of each of the ECFIM schemes being compared.

\subsection{Unpreconditioned Jacobian-Free Newton-Krylov}
The Jacobian Free (JF) strategy for the Newton Krylov (NK) approach is extensively described in the literature \cite{knoll2004jacobian}. In essence, the Newton method is used to compute the iterations of the solution of a non-linear set of  equations. Each Newton iteration requires the solution of the linear equation represented by a Jacobian matrix. The NK uses Krylov-space methods for solving the linear problem (we use GMRES in the present study \cite{saad1986gmres}). The JF implementation avoids computing the Jacobian by instead approximating numerically its product by a Krylov-space vector. The JFNK has the great advantage of lending itself to being used as a black box. We use in the present study the textbook implementation for MATLAB described in Ref. \cite{kelley}.

We define the unknown state vector made of the particle velocity and variation of scalar potential $x_k=\{\{\overline{\bfv}_p\}, \{\delta \phi_g\}\}$ and we refer to it as the Krylov vector of unknowns.  From it any other quantity can be constructed: 
\begin{itemize}
\item the new particle position using the old position and $\overline{\bfv}_p$ from the Krylov vector
\item the current and its divergence by particle interpolation, 
\item the electric field from the potential $\delta \phi_g$.
\end{itemize}
The Krylov vector summarises the complete state of the system at the time centered level $t+\Delta t/2$. 
The JFNK method requires the definition of a residual that is computed based on the guesses of the Krylov vector provided by the successive iterations of the Newton method. The user only needs  to specify the residual. In this case, the residual is given by the Newton velocity equation and by the Poisson equation. All other equations required are just ancillary equations that do not directly produce a residual, but only temporary quantities needed to compute the two main residual equations. In Matlab, this is implemented as:

\begin{lstlisting}[caption=Skeleton code for residue of the unpreconditioned JFNK implementation, label=unprec-JFNK]
% Residue of the equation of motion 
res(1:N) = xkrylov(1:N) - v0 - 0.25*mat*QM*(2*E0(1:NG)+dE(1:NG))*DT;
% Residue of the Poisson equation
res(N+2:N+NG-1) = ( dphi(1:NG-2)+dphi(3:NG)-2*dphi(2:NG-1) )/dx^2 ;
% Apply periodic boundary conditions
res(N+1) = ( dphi(NG)+dphi(2)-2*dphi(1) )/dx^2 ;
res(N+NG) = ( dphi(NG-1)+dphi(1)-2*dphi(NG) )/dx^2 ;
% Call particle projection
J = project_particles;
divJ = divergence(J);
% Add sources produced by particle projection
res(N+1:N+NG) = res(N+1:N+NG) - divJ*DT;
\end{lstlisting}
where \texttt{xkrylov}  is the Krylov vector with the first \texttt{N} positions being the particle velocities and the last \texttt{NG} being the potential variations ( \texttt{N} the number of particles and  \texttt{NG} the number of grid points). The current and its divergence are computed by particle interpolation (see  the full code reported in Appendix \ref{ECFIM-unprec} for details). 
With this residual definition, JFNK can solve the set of coupled non-linear equations with a simple call:
\begin{lstlisting}
[sol, it_hist, ierr] = nsolgm(xkrylov,'residue',tol,parms);
\end{lstlisting}
\noindent where \texttt{residue} if the function evolution of the residual. Tolerance and parameters to control the iteration procedure are chosen with the guidelines illustrated in Ref. \cite{kelley}.
Obviously, it is hard to imagine anything simpler, but this formal simplicity is paid for by the computational complexity offloaded to the JFNK method. In particular this has two downsides:
\begin{enumerate}
\item The Krylov vector is large, with dimension N+NG that includes all particles and all cell points. As a consequence, all the internally constructed vectors and matrices of the JFNK tool will have to deal with this large dimension. The JFNK additionally uses GMRES for solving the linear Jacobian problem with the additional need to store multiple Krylov subspace vectors for the orthogonalization. Hence, the memory requirement becomes heavy.
\item The method is completely physics-agnostic, it does not use anything in our knowledge of the physics. For example, we know each particle is formally independent of all others, in the sense that the partial derivative of one particle residual taken with respect to the other particle's velocities is zero. But this information is discovered by JFNK by trying to compute the corresponding elements of the Jacobian. This waste of knowledge is paid in terms of computational wasted effort due to a higher number of Newton and Krylov iterations.
\end{enumerate}
Nevertheless, the know-nothing approach is often the best initial approach, and is used here for sake of comparison, expecting all other methods to perform better.

\subsection{Particle Hiding Preconditioning } 
An obvious strategy that avoids the first of the pitfalls above is to realise that, once the fields are guessed by the Newton iteration, the particles can be moved without being part of the Newton iteration itself. In fact, moving particles is an independent  task that can be done particle by particle. In the particle hiding strategy, the Krylov vector only has the field unknowns:   $x_k=\{\delta \phi_g\}$. 
At each residual evaluation, the particles are updated from the  position and velocity of the previous time step to the new values determined by the fields present in the current Newton iteration. When the fields have converged to their value, the last particle motion done for the last residual evaluation gives the final particle values consistent with the converged fields. 
The new residual evaluation is then just:
\begin{lstlisting}[caption=Skeleton code for residue JFNK implementation with particle hiding, label=unprec-JFNK]
% Particle Hiding Preconditioner
particle_mover;
J = project_particles;
divJ = divergence(J);
% Residual Poisson
res(2:NG-1) = ( dphi(1:NG-2)+dphi(3:NG)-2*dphi(2:NG-1) )/dx^2 ;
res(1) = ( dphi(NG)+dphi(2)-2*dphi(1) )/dx^2 ;
res(NG) = ( dphi(NG-1)+dphi(1)-2*dphi(NG) )/dx^2 ;
res(1:NG) = res(1:NG) - divJ*DT;
\end{lstlisting}
where hidden in the use of \texttt{divJ}  are the operations of moving the particles in the fields provided by the most updated Newton iteration, interpolating them to the grid to compute the current and taking the divergence. See Appendix \ref{ECFIM-phiding} for a full listing of the residual evaluation. 

The advantage is clear: the Krylov vector has just \texttt{NG} elements. As a consequence, all internal storage for the JFNK package is  drastically reduced in size. But the downside is equally clear:  now at each residual evaluation the particles need to be moved. This drawback, however, is not as serious as it sounds. First, the residual evaluation of the unpreconditioned case also essentially requires moving the particles so the difference in cost is not as severe as one might cursorily suspect. Second, decoupling particle motion and NK iteration has the advantage of giving more options on how to move the particles, such as for example subcycling the particle motion \cite{chen2011energy} or using multiscale ODE solvers for the equations of motion \cite{Brackbill:1985}. Another advantage of this approach is that having decoupled fields and particles, the two parts can be computed in different hardware parts, for example the particles can be moved in boosters (GPU or MIC) while the NK iteration is handled by CPUs \cite{mallon2012scalability,chen2012efficient}.

In the present case we do not use Cluster-Booster architectures and simply move the particles using the same algorithm as in Eq. (\ref{eq:ECPICmover_av}) and solving the equation of motion  independently for each particle. This choice is made for simplicity but the reader should keep in mind that a skilled implementation on a hybrid architecture can improve the performance of this approach considerably  \cite{chen2012efficient}.

\subsection{Field Hiding Preconditioning}

A complementary choice allows to remove the second pitfall of the unpreconditioned JFNK implementation, but not the first: field hiding. To our knowledge this approach has not yet been proposed in the literature, perhaps because at first sight it seems unpromising as still suffering from the first pitfall (large Krylov vectors).  However, the method of using the particle moments to precondition the NK iteration reaches similar goals \cite{chen2014fluid,taitano2013development}.

In the field hiding approach, the Krylov vector has only the particles and not the fields, the opposite of the particle hiding approach: $x_k=\{\overline{\bfx}_p\}$. The rationale for this is that we can always solve the linear field equations without needing any Newton iteration if we know the new updated particle positions. During the residual evaluation, we can  assume as known the particle velocity (and consequently the particle position), provided by the most updated Newton iteration. Based on that, solving the field equations is a relatively simple exercise of linear field update as is done in many computational electrodynamics packages \cite{taflove}. 
The non-linearity in implicit  PIC methods comes from the coupling of fields and particles: here instead we can access the particles from the Newton iteration guess and the field solution then translates into a simple linear problem. In the present 1D case this can be trivially done as a 3-diagonal solver, whilst other efficient direct or multigrid or Fourier space solvers are available in 2D and 3D \cite{taflove}.
In this case, the residual evaluation is then:
\begin{lstlisting}[caption=Skeleton code for residue JFNK implementation with field hiding, label=unprec-JFNK]
% advancing  x at n+1/2 time level based on the Newton iteration of the particle velocity
x_average = x0 + xkrylov(1:N)*DT/2;
% applying BC
out=(x_average<0); x_average(out)=x_average(out)+L;
out=(x_average>=L);x_average(out)=x_average(out)-L;
% Field Hiding Preconditioner
J = project_particles;
divJ = divergence(J);
dE = direct_field_solver;
% residual for the average velocity
res(1:N) = xkrylov(1:N) - v0 - 0.25*mat*QM*(2*E0(1:NG)+dE(1:NG))*DT;\end{lstlisting}
where hidden in the use of \texttt{dE} there is computing the field equations (see \ref{ECFIM-fieldhiding} for the full MATLAB listing). 
The downside of the field hiding approach is obviously that the Krylov subspace still has the dimension of the number of particles, with all memory requirements that it implies. But the powerful advantage is that the fields are the main cause for which the whole system needs to be iterated.  The coupled system of particles and fields are coupled because the fields couple the particles: particles do not directly interact with each other in PIC methods, their interaction is mediated by the fields. In the unpreconditioned method this means that the derivative of one particle residual with respect to another particle is zero. However, the derivatives of the field residual with respect to all particles are non-zero and the derivatives of the particle residual with respect to the fields are also non-zero. The real coupling, 
as well as the real communication in parallel implementations, is due to the field part. 
In using the field hiding preconditioning, we have taken away from the Newton iteration basically all couplings, the coupling via the fields is handled by the linear solver for the field equations. As a result, we expect the Newton iteration to converge in very few iterations, a fact confirmed by the results.

\subsection{Solution with Direct Newton-Schwarz}
As mentioned above, the particle hiding strategy resolves the first pitfall of the unpreconditioned JFNK (i.e. it reduces drastically the memory requirement), while the field hiding  resolves  the second (the absence of physical insight into the  solution strategy).  But neither of them resolves both at once.  With  the Direct  Newton-Schwarz (DNS) method both pitfalls are eliminated. 
The DNS method  is still  based on the field hiding approach, i.e. it uses  an optimised  solver  for the field equations but it also eliminates the need for a large-dimensional Krylov subspace. As noted above, the  Jacobian elements where cross particle  derivatives are computed are zero, so why use Krylov subspace methods at all? 
We can follow a different path. We can obtain the Newton update using a Schwarz decomposition \cite{cai1998parallel} of the Jacobian equation of the field hiding JFNK method. We compute analytically the Jacobian of the sub-problem of the residual equation for just one particle.  We then obtain the new Newton iteration inverting this simple Jacobian, i.e. just a scalar in the present case (or 3x3 matrix in 3D): $\frac{\partial Res(v_p)}{\partial v_p}$.  Based on the new  Newton iteration for each particle  velocity, we can proceed exactly as in the field hiding preconditioner to compute the  fields. 
There are two fundamental differences between  JFNK with field hiding and DNS. First, there is no Krylov solver used, so there is no memory requirement for it, the particle equations of motion are computed by the Newton method with a  particle-by-particle Jacobian computed analytically. Second, the  physics coupling carried by the fields is still solved directly using a linear field solver. 
The   DNS method then combines  the separate advantages of the two hiding preconditioners without sharing any of the pitfalls. But  is it perfect? No. And the reason is that in Schwarz-decomposing the Jacobian we cheated on the math.  It is true that the full Jacobian of the particles and fields  of the unpreconditioned method has zero in all elements when the residual of particles is  derived with respect to the each others' velocities.
But  when the residuals are computed only for the particles and the fields are hidden, the dependence of the electric field on the positions and velocity (via the current) of all other particles comes back. This effect is present in the derivative of the field residual with respect to the particle velocities in the full  Jacobian of the unpreconditioned case. But when the residual is computed only for the particles, it comes  to coupling directly all particles with all other particles (or at least a larger set of them).  Using the Jacobian decomposition particle-by-particle we neglect this coupling and we run the risk that more iterations will be needed. Indeed,  if the  decomposition of the Jacobian were exact, the DNS would converge in one iteration,  while instead the results below show that some extra iterations are necessary. But  their number is small because those couplings are already almost completely addressed by using the direct field solver of the field hiding strategy.
The script for the DNS is:
\begin{lstlisting}[caption=Skeleton code for DNS, label=unprec-JFNK]
% calculate the x at n+1/2 time level
x_average = x0 + v_average(1:N)*DT/2;
% apply BC
out=(x_average<0); x_average(out)=x_average(out)+L;
out=(x_average>=L);x_average(out)=x_average(out)-L;
%  compute Current and solve Field Equation  directly
J = project_particles;
divJ = divergence(J);
dE = direct_field_solver;
% apply direct Newton method assuming Schwarz decomposed  Jacobian
fraz=[(fraz1);1-fraz1];	
mat=sparse(p,g,fraz,N,NG);
frazder=[-ones(size(fraz1)) ones(size(fraz1))]/dx;
matder=sparse(p,g,fraz_der,N,NG);
res(1:N) = xkrylov(1:N) - v0 - 0.25*mat*QM* 2*E0(1:NG)+dE(1:NG))*DT;
schwartz_jacobian(1:N) = -1 + 0.125*mat_der*QM*(2*E0(1:NG)+dE(1:NG))*DT.^2;
v_average_new(1:N) = v_average(1:N)+ res(1:N)./schwartz_jacobian(1:N);
% compute norm of the iteration error and prepare for next iteration
error=norm(v_average_new-v_average);
v_average(1:N) = v_average_new;
\end{lstlisting}
Note the need to compute not only the interpolation function (\texttt{mat}) but also its derivative  (\texttt{matder}).
The approach above  is almost as simple as a  Picard iteration  \cite{taitano2013development}, including however the information from the Jacobian. This approach looks very trivial in  one velocity dimension without magnetic fields,  but it becomes a little more involved  for the general three  velocity  dimensions and  in presence of a magnetic field,  or in the relativistic case. Still even in those cases it is just a $3\times 3$ Jacobian that can be computed and inverted analytically. The script above, supplemented by initial conditions and a criterion to stop the iteration, is all that is needed. This method is by far the simplest of all considered in the present paper, a bonus in terms of  implementation on large hybrid parallel computers. 
The DNS turns out to be also the fastest of all the ECFIM analyzed here, as well as the one that requires the least memory.

\section{Results} \label{sec:results}
\begin{figure}[h!]
\centering
\includegraphics[width=\columnwidth, angle =0]{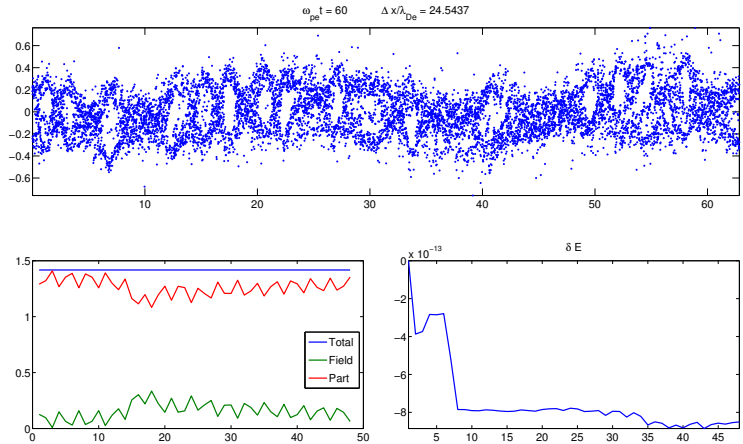}
\caption{Example of the output obtained from running the unpreconditioned version of the code proposed. The upper panel represents the electrons phase-space, upon 
which the clear features of a two-stream instability can be appreciated. The lower panels represent the energy evolution (left panel) and its relative error 
respect to the initial value.
Specifics of this simulation are: $10^{-14}$ JFNK tolerance,  $10000$ particles and $256$ cells.} 
\label{fig:two}
\end{figure}

To evaluate the performance of the implementations discussed above,  we present a parametric study organized in four tests: 
\begin{itemize}
\item the first test measures the performance with a different time step amplitude, 
\item the second test measures  the performance with a different tolerance of the Newton method, 
\item the third test measures the performance with a different resolution (i.e. cell size), 
\item the fourth test measures the performance with a different number of particles.
\end{itemize}
In particular, we set the analysis around an electron two stream instability in a neutral plasma featuring cold and fixed ions, with an  electrons drift velocity $V_e = 0.2\,c$. 
The study of the two-stream instability is one of the first initial approaches normally taken to test a new plasma-related code. Explanations on how this
instability works can be found in plasma-physics books, such as in \cite{birdsall-langdon}.
Particles are distributed in a 1D domain with size $L=20\pi\,c/{\omega}_{pe}$. Furthermore, periodic boundary conditions are assumed and the system is let evolve for 200 cycles. To make a clearer illustration, since each parametric study performs the change of a single parameter, it is convenient to establish the parameters of the reference set-up:  the tolerance of the Newton and GMRES algorithms are equal to $10^{-10}$ (where present), ${\omega}_{pe}\Delta t = 0.1$, $N_p = 10^5$ and $N_g = 64$. Anywhere not  mentioned  otherwise the corresponding parameter equals the one in the reference set-up.
All tests include the comparison of 5 different methods:
\begin{enumerate}
\item ECFIM with no preconditioner,
\item ECFIM with particle-hiding preconditioner,
\item ECFIM with field-hiding preconditioner,
\item the Direct Newton Schwarz (DNS) method,
\item the semi-implicit code ECSIM.
\end{enumerate}
An example of the output obtained from running the unpreconditioned version of the code is given in Fig. \ref{fig:two}, 
where signatures of the two stream instability are clearly visible in
the particles phase-space (upper panel), together with the energy evolution and its relative error respect to the initial value (lower panels).
This test featured a non-linear iteration tolerance of $10^{-14}$,  $10000$ particles and $256$ cells.

\subsection{Test 1 - Different Time Step} \label{sec:timestep}

\begin{figure}[h!]
\centering
\includegraphics[width=\columnwidth, angle =0]{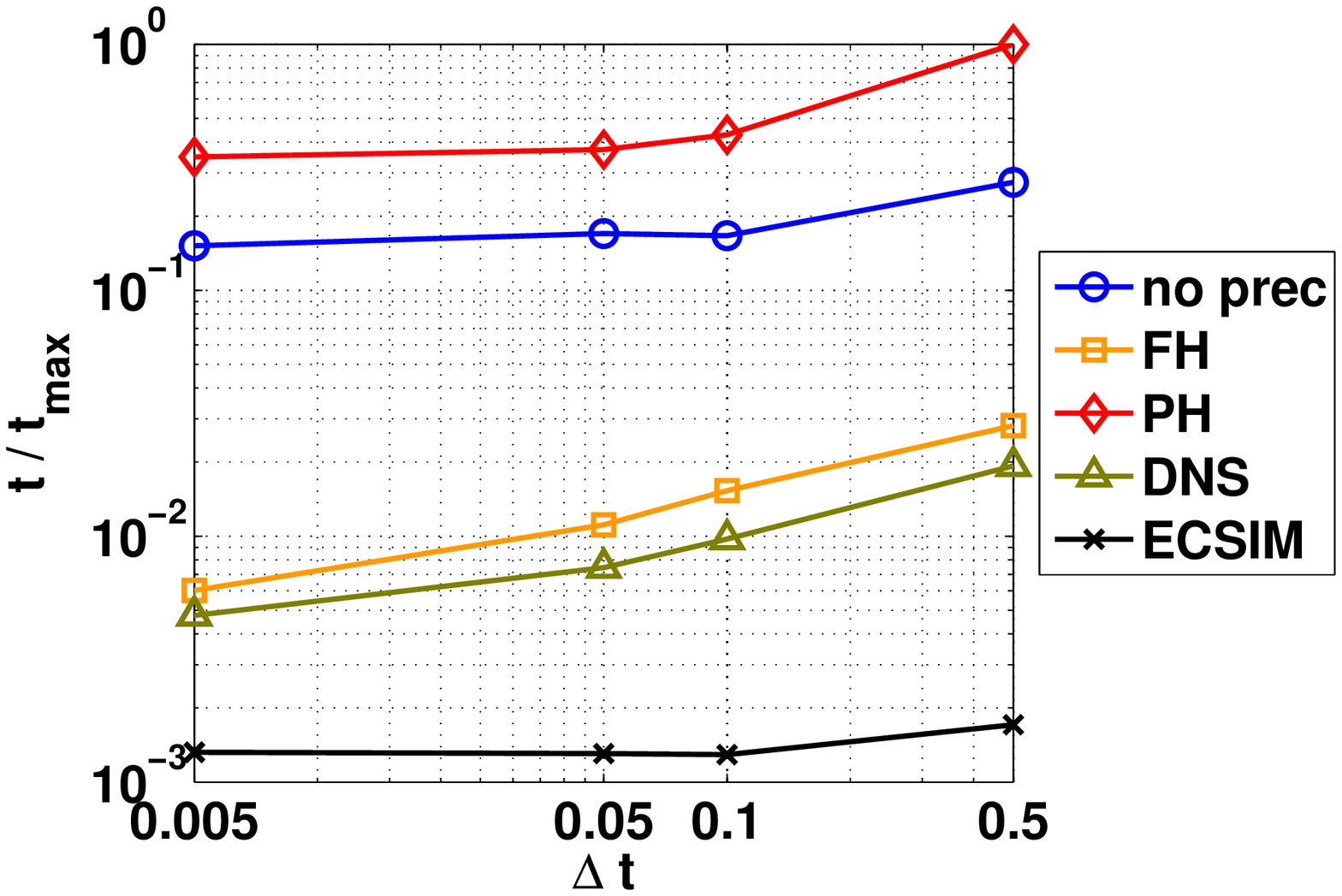}
\caption{Dependence of the relative run time per cycle as a function of the time step, namely ${\omega}_{pe} \Delta t = 0.005,\,0.05,\,0.1,\,0.5$.  The run time is normalized to the maximum time consumed over all the simulations considered.
The following methods are shown: ECFIM with no preconditioner (blue circles), ECFIM with particle-hiding (red diamond), ECFIM with field-hiding (orange squares), ECSIM (black Xs) and direct newton solver (green triangles).} 
\label{fig:dt}
\end{figure}

\noindent From the computational point of view, the time step size may influence the convergence in time dependent problems. Therefore, this section is devoted to analyse the behaviour of the algorithms at different time step.
In Fig. \ref{fig:dt} we plot the time considered by the code with different $\Delta t$. Some interesting considerations can be made.
First of all, we notice the ECSIM approach to be the fastest. The ECSIM method requires the computation of a mass matrix, adding extra costs, but it does not use any Newton iteration. Specifically, the cost per cycle in ECSIM is nearly independent of the time step used: this finding is not surprising given the absence of Newton iterations, therefore  the cost per cycle is expected to be more constant. However, ECSIM still requires the solution of a linear set of equations, leading to a slight hump in the curve for low $\Delta t$. Notice that the cost  of ECSIM is nearly the same regardless the solver used, either the MATLAB direct solver or GMRES. 

After ECSIM, the following best performing approaches are the DNS and the field hiding (FH). The DNS method is faster by a visible factor, while the performance ratio between the two  remains the same as $\Delta t$ changes. In both the cases the number of Newton iterations increases with $\Delta t$ and so does the cost per cycle.

Finally, the particle hiding approach and the unpreconditioned ECFIM perform the worst. In fact, the case with the particle hiding is even slower than when no preconditioning is set. Neither of the methods includes a preconditioning capable to capture the field-mediated couplings, leading therefore to a higher number of residual evaluations. In the particular case of the particle hiding, the particle mover does not employ any Eisenstat-Walker inexact convergence \cite{kelley}.  The inexact convergence uses a smart approach to relax the tolerance for the linear GMRES iterations when the non linear Newton convergence is still far from being achieved. This approach limits the cost of the unpreconditioned ECFIM. When the particle hiding is used, the mover moves the particles with a precision requiring the minimum tolerance allowed, leading  to a larger overall computational effort.

\subsection{Test 2 - Different Tolerance of the Newton's Method} \label{sec:tolerance}

\begin{figure}[h!]
\centering
\includegraphics[width=\columnwidth, angle =0]{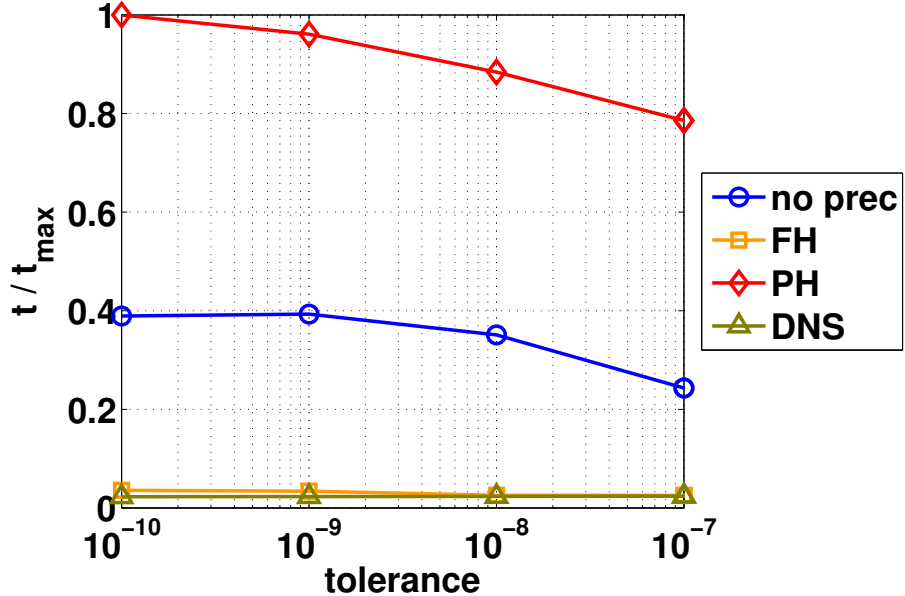}
\caption{Dependence of the relative run time per cycle as a function of tolerance value for the non-linear iterative solver, namely $10^{-7},\,10^{-8},\,10^{-9},\,10^{-10}$.  The run time is normalized to the maximum time consumed over all the simulations considered.
The following methods are shown: ECFIM with no preconditioner (blue circles), ECFIM with particle-hiding (red diamond), ECFIM with field-hiding (orange squares) and direct newton solver (green triangle).} 
\label{fig:iter_tol_plot}
\end{figure}

\begin{figure}[h!]
\centering
\includegraphics[width=\columnwidth, angle =0]{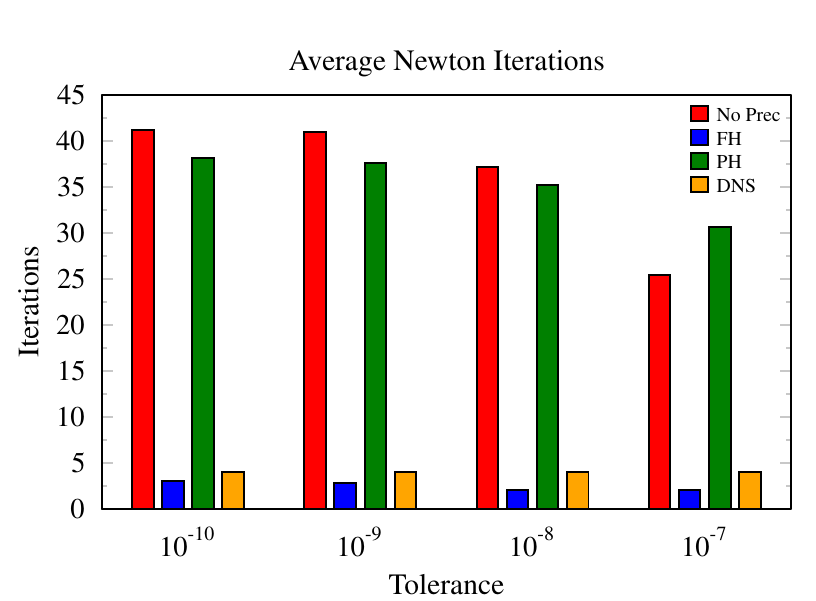}
\caption{Average number of Newton iterations performed in the non-linear iterative cycle necessary to the methods presented (red bar for ECFIM with no preconditioner, green bar for ECFIM with particle-hiding, blue bar for ECFIM with field-hiding and orange bar for the direct newton solver). Notice that the ECSIM does not require any iterative cycle by construction.} 
\label{fig:iter_tol}
\end{figure}

This test shows the influence of different   convergence tolerances of the Newton's iterative method on the simulation performance.
The time step is fixed to ${\omega}_{pi} \Delta t=0.1 $, and the other parameters are the same as the standard case. 
 The number of iterations of the Newton solver depends on the tolerance considered. Figures \ref{fig:iter_tol_plot} and \ref{fig:iter_tol} show a certain independence of the chosen tolerance only for the DNS and FH methods, while the other preconditioners reach a saturation point for a tolerance around $10^{-9}$.
In fact we observe an effective improvement in terms of the global performance with these two latter field-hiding preconditioners. In fact, both the field-hiding and DNS drastically reduce the average number of non linear iterations given its capability of capturing better the particle-field couplings with its strategy, leading then to a more rapid Newton convergence. But the field-hiding leads to the smallest number of iterations. This finding is not surprising as the DNS approximates the Jacobian with the Schwarz decomposition, requiring hence more iterations than the case with full a Jacobian evaluation. However, note that each DNS iteration is less computational intensive than that of the field-hiding, which can be explained as with the Jacobian being inverted analytically.

\subsection{Test 3 - Different Mesh Size} \label{sec:meshsize}

\begin{figure}[h!]
\centering
\includegraphics[width=\columnwidth, angle =0]{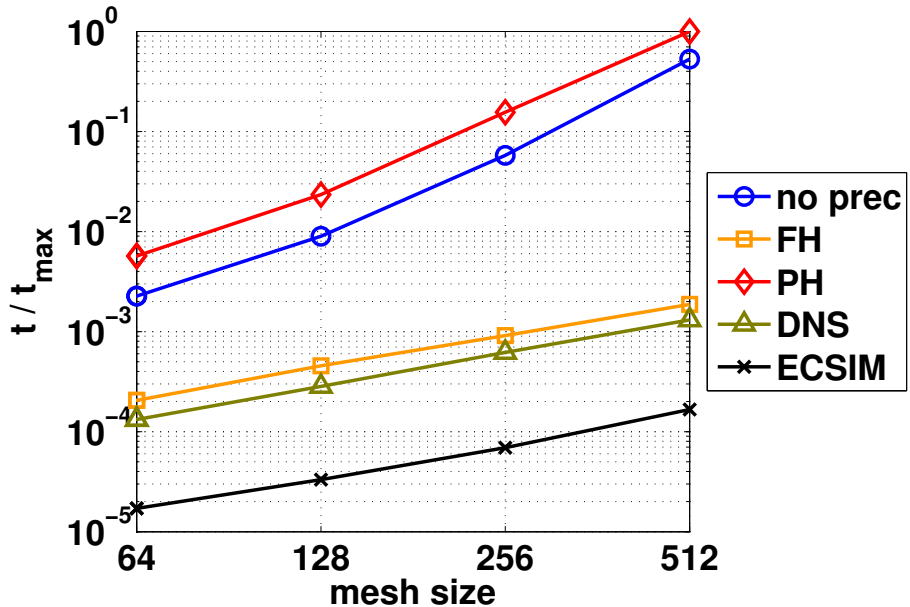}
\caption{Dependence of the relative run time per cycle as a function of the domain resolution, namely $x = 64,\,128,\,256,\,512$.  The run time is normalized to the maximum time consumed over all the simulations considered. 
The following methods are shown: ECFIM with no preconditioner (blue circles), ECFIM with particle-hiding (red diamond), ECFIM with field-hiding (orange square), ECSIM (black Xs) and direct newton solver (green triangles).} 
\label{fig:mesh}
\end{figure}

Of  critical interest for practical applications is the performance behaviour when varying the mesh size. Obviously production runs will need large multidimensional grids, so it is important to see how the different implementations scale as the number of cells increases. This test is set such that the mesh size increases as the ratio $N_p/N_{g}$ is kept constant, where $N_g$ and $N_{p}$ are respectively the number of cells and particles (see Table \ref{table:mesh}).
\begin{table}[h!]
\centering
\caption{Summary Table of the setup considered to study the methods performance under a different resolution. Runs are set so that the ratio $N_p/N_{g}$ be kept constant. Therefore, the number of particles in the system changes accordingly.}
\label{table:mesh}
\begin{tabular}{l|lllll}
$N_g$ & 64  & 128 & 256 & 512 \\ \hline
$N_p$   & 10K & 20K & 40K & 80K 
\end{tabular}
\end{table}
The improvement in the statistic implies an increase in computational effort, which may have a important effect on the couplings in the Maxwell's equation.  
Figure  \ref{fig:mesh} reports the relative timing of each method, normalised to the longest run among all tested.  As in previous tests, field hiding JFNK and DNS show a better scaling, comparable with that of the ECSIM method. The other two cases (unpreconditioned and  particle-hiding) show a less desirable scaling, leading to a faster growth in computational time as the simulation size increases.

To analyse the performance deterioration cause of the unpreconditioned and particle hiding approach, table \ref{tab:iter_stat} shows the  number of newton iterations for each method at different grid sizes.   While the preconditioners field hiding and DNS require only an average of three Newton iterations, the particle hiding and the full coupled algorithm require a fast growing number of iterations. Concretely, the table \ref{tab:iter_stat} shows that the non-linear solver may be a bottleneck if it is not carefully preconditioned. Furthermore, the raise in Newton iteration is a clear evidence of the the different asymptotic behaviour in figure \ref{fig:mesh}. 

\begin{table}[]
\centering
\caption{Average Newton iteration performed in the study on the mesh size. ECFIM with no preconditioner (No Prec), ECFIM with field-hiding (FH), ECFIM with particle-hiding (PH) and the direct Newton solver (DNS) are compared. Notice that the ECSIM does not require any iterative cycle by design.}
\label{tab:iter_stat}
\begin{tabular}{l|llll}
Mesh size & No prec                   & FH   & PH     & DNS  \\ \hline
64        & 41.33                     & 2.96 & 38.315 & 3.98 \\
128       & \multicolumn{1}{c}{70.41} & 2.98 & 69.71  & 3.99 \\
256       & 204.10                    & 2.99 & 210.20 & 4.0  \\
512       & 843.27                    & 2.99 & 612.64 & 4.0 
\end{tabular}
\end{table}

The lesson of this test is that the couplings across the domain carried by the fields need to be preconditioned away from  the non linear iteration or the costs of the simulation balloon as the simulation size increases. 

\subsection{Test 4 - Different Number of Particles} \label{sec:numb_particles}

The final test proposed here regards the influence of the number of particles on the simulation performance, when the grid is not also increased. The interest of this test is that as the number of particles is increased the noise of the PIC simulation decreases. We can then test the effect of noise on convergence. 

\begin{figure}[h!]
\centering
\includegraphics[width=\columnwidth, angle =0]{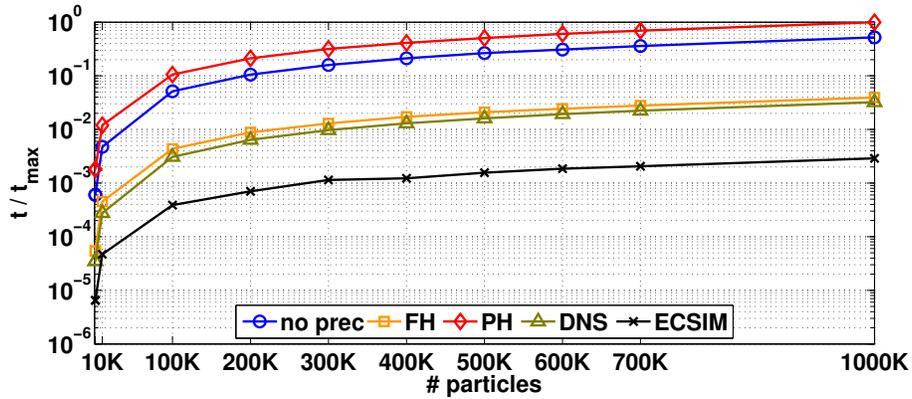}
\caption{Dependence of the relative run time per cycle as a function of the number of particles within the domain.
A system with, respectively, $10^3,\,10^4,\,10^5,\,2 \cdot 10^5,\,4 \cdot 10^5,\,8 \cdot 10^5 \text{ and } 10^{6}$ has been considered.  The run time is normalized to the maximum time consumed over all the simulations considered.
The following methods are shown: ECFIM with no preconditioner (blue circles), ECFIM with particle-hiding (red diamond), ECFIM with field-hiding (orange square), ECSIM (black Xs) and direct newton solver (green triangles).} 
\label{fig:part}
\end{figure}

Unlike the previous plots, results in Fig. \ref{fig:part} are shown in a semi-logarithmic scale for a better readability. We chose a broad range of cases, spanning from that with a very few particles simulated (i.e. $1000$) through the case with a great number of particles in the system (i.e. $10^{6}$).  
As could be expected, the reduction in noise at higher number of particles leads to a saturation of the computational cost. This feature seems to be of particular importance given the kinetic nature of these methods, allowing for the usage of a large number of particles without dramatically increasing the computational time.
The best performance is again achieved by   ECSIM. Also, the two methods sharing the field-hiding method show a very similar behaviour featuring nearly the same execution time. This was somehow expected given their common original approach, even though the DNS method seems to be not particularly more affective than the original field-hiding preconditioner. This is in line with the the concept of DNS being conceived for a memory use improvement rather than  execution time. 

\section{Conclusions} \label{sec:conclusions}

In this paper we presented a set of different strategies to improve the computing  performance of fully implicit Particle-in-Cell (PIC) codes. Unlike semi-implicit PICs, which make
use of a linear approximation to improve performance, we consider here no approximations of the original formulation and focus only on advanced non-linear solvers that still produce the same exact result.
Due to the heavy CPU and memory consumption requested by the  iteration of the entire non-linear system, the fully implicit approach has long been considered expendable over the semi-implicit counterpart. This concern, however, is largely eliminated when preconditioners are used.
In this work we proved that the computation with fully implicit PICs can be significantly improved whether a careful application of ad-hoc preconditioning methods be applied. The semi-implicit method remains the fastest, but the fully implicit methods come now close second.

The first preconditioner method considered here is found in previous literature \cite{markidis2011energy} and named particle-hiding after its approach of segregating  the particle mover from the field solver in the non-linear JFNK iteration. Particles are still continuously updated within the iterative cycle using the fields information, but now the particles velocity is no longer considered as part of the residual convergence, consequently leading to a lower number of iterations and a strong reduction of the 
memory demanded due to the non storage of the particles information. The downside is that for each residual evaluation, all particles still need to be moved, 
so that this approach still relies on the full mutual coupling between particles and fields, which conducts to further computational costs.

We have then proposed a further improvement based upon the fact that are actually the fields being the  coupling  source in the system rather than the particles. Particles do not directly interact with each other, their interaction is provided by the field. Mathematically, the fields are coupled with each particles while the particles just need the updated field information to compute their new position and velocity.
The approach is based on a complementary variant of the previous method, and is named field-hiding after its application to the field information instead of particles'. It is now the field 
solver to be segregated out of the residual computation, so that the iteration is only performed over the particles velocity. Having removed all direct couplings from the JFNK calculation and having relegated all field couplings to the preconditioner, this approach is shown to perform faster then the previous one. However, the memory consumption downside is yet present, as 
the method requires a memory allocation as large as the number of particles considered in the system, which may become very heavy for real-system simulations.

In order to further relax this issue, a  different preconditioner based on the field-hiding approach has been presented. Given the independence of the particles from each other, the Jacobian matrix appears to have all the cross-over derivatives equal to zero. We can then  consider to eliminate the use of the Krylov space and resolve the linear part of the Newton iteration using a Schwarz decomposition. The JFNK is then no longer needed, each particles is solved independently via a direct newton iteration based on the Schwarz decomposed Jacobian computed and inverted analytically particle by particle.  In doing so, we combine  the two main advantages of the two previous preconditioners: 1) not  having to deal with a large dimensional array and 2) not dealing with the main couplings directly inside the non-linear iterative loop. 

This latter approach is shown to lead to a lower number of iterations required in the non-linear solver and to the fastest CPU performance. From the tests presented, we can in fact confirm that the Direct-Newton-Schwarz (DNS) method is the fastest to converge. The DNS also eliminates the need to reserve memory for the particle iterative solution as each particle is solved independently and directly. DNS comes second only to the semi-implicit ECSIM method, which yet relies on a completely different approximation approach. 

In conclusion, we observe how the field-hiding preconditioner approach is reported here as the most promising, especially when coupled with a DNS formulation, avoiding Krylov subspace solvers altogether. In fact, all the tests with different number of particles, different mesh, different Newton iteration tolerance and different time step have shown that the field-hiding approach, and especially its DNS implementation, is the best fully non-linear system. While the ECsim semi-implicit approach is still the fastest, we note that its extension to relativistic system does not appear to be straight-forward, while that of the DNS fully implicit method is.  

\appendix
\section{MATLAB Listing of the unpreconditioned residual}\label{ECFIM-unprec}
\begin{lstlisting}[frame=none]
% calculate the x at n+1/2 time level
x_average = x0 + xkrylov(1:N)*DT/2;

% apply periodic particle boundary conditions
out=(x_average<0); x_average(out)=x_average(out)+L;
out=(x_average>=L);x_average(out)=x_average(out)-L;

% calculate the interpolation functions
p=1:N;p=[p p];
g1=floor(x_average/dx)+1;     
g=[g1;g1+1];
fraz1=1-abs(x_average(1:N)/dx-g1+1); 	
out=(g<1);g(out)=g(out) + NG;
out=(g>NG);g(out)=g(out)- NG;

% calculate the current J with periodicity
fraz=[(fraz1).*xkrylov(1:N);(1-fraz1).*xkrylov(1:N)];	
mat=sparse(p,g,fraz,N,NG);
J = full((Q/dx)*sum(mat))';
J(NG+1)=J(1); 
% compute the divergence of J
divJ=zeros(NG,1);
divJ(1:NG)=(J(2:NG+1)-J(1:NG))/dx;

% residual of the Poisson Equation
res = zeros(N+NG,1);
res(N+2:N+NG-1) = ( dphi(1:NG-2)+dphi(3:NG)-2*dphi(2:NG-1) )/dx^2 ;
% apply periodic boundary conditions
res(N+1) = ( dphi(NG)+dphi(2)-2*dphi(1) )/dx^2 ;
res(N+NG) = ( dphi(NG-1)+dphi(1)-2*dphi(NG) )/dx^2 ;
% add the source from the particles
res(N+1:N+NG) = res(N+1:N+NG) - divJ*DT;

% compute the Electric field
dphi= xkrylov(N+1:N+NG);
dE=zeros(NG+1,1);
dE(2:NG) = -(dphi(2:NG)-dphi(1:NG-1))/dx;
dE(1) = -(dphi(1)-dphi(NG))/dx;
dE(NG+1) = dE(1);

% residual of the equations of motion electrons 
fraz=[(fraz1);1-fraz1];	
mat=sparse(p,g,fraz,N,NG);
res(1:N) = xkrylov(1:N) - v0 - 0.25*mat*QM*(2*E0(1:NG)+dE(1:NG))*DT;
\end{lstlisting}

We refer the reader to  \cite{lapenta-book:2017} for any unclear aspect of the  program. Especially the implementation of the interpolation function using MATLAB matrices is not trivial, and all details are fully explained in the reference.

\section{MATLAB Listing of the residual with particle  hiding}\label{ECFIM-phiding}
\begin{lstlisting}[frame=none]
dphi= xkrylov(1:NG);

% Mover for the particles hidden from the main NK iteration
v_average = particle_mover(E0, dphi);
x_average = x0 + v_average*DT/2;   
out=(x_average<0); x_average(out)=x_average(out)+L;
out=(x_average>=L);x_average(out)=x_average(out)-L;

% calculate the interpolation functions
p=1:N;p=[p p];
g1=floor(x_average/dx)+1;     
g=[g1;g1+1];
fraz1=1-abs(x_average(1:N)/dx-g1+1); 	
out=(g<1);g(out)=g(out) + NG;
out=(g>NG);g(out)=g(out)- NG;

% calculate the current J
fraz=[(fraz1).*v_average(1:N);(1-fraz1).*v_average(1:N)];	
mat=sparse(p,g,fraz,N,NG);
J = full((Q/dx)*sum(mat))';
J(NG+1)=J(1);

divJ=zeros(NG,1);
divJ(1:NG)=(J(2:NG+1)-J(1:NG))/dx;

res = zeros(NG,1);

dE=zeros(NG+1,1);
dE(2:NG) = -(dphi(2:NG)-dphi(1:NG-1))/dx;
dE(1) = -(dphi(1)-dphi(NG))/dx;
dE(NG+1) = dE(1);

% Residual Poisson
res(2:NG-1) = ( dphi(1:NG-2)+dphi(3:NG)-2*dphi(2:NG-1) )/dx^2 ;
res(1) = ( dphi(NG)+dphi(2)-2*dphi(1) )/dx^2 ;
res(NG) = ( dphi(NG-1)+dphi(1)-2*dphi(NG) )/dx^2 ;
res(1:NG) = res(1:NG) - divJ*DT;

\end{lstlisting}

\section{MATLAB Listing of the residual with field  hiding}\label{ECFIM-fieldhiding}
\begin{lstlisting}[frame=none]
% calculate the x at n+1/2 time level
x_average = x0 + xkrylov(1:N)*DT/2;
out=(x_average<0); x_average(out)=x_average(out)+L;
out=(x_average>=L);x_average(out)=x_average(out)-L;

% calculate the interpolation functions
p=1:N;p=[p p];
g1=floor(x_average/dx)+1;     
g=[g1;g1+1];
fraz1=1-abs(x_average(1:N)/dx-g1+1); 	
out=(g<1);g(out)=g(out) + NG;
out=(g>NG);g(out)=g(out)- NG;

% calculate the J
fraz=[(fraz1).*xkrylov(1:N);(1-fraz1).*xkrylov(1:N)];	
mat=sparse(p,g,fraz,N,NG);
J = full((Q/dx)*sum(mat))';
J(NG+1)=J(1);

divJ=zeros(NG,1);
divJ(1:NG)=(J(2:NG+1)-J(1:NG))/dx;

res = zeros(N,1);

% Linear Field Solver - Direct solver implementation
un=ones(NG-1,1);
Poisson=spdiags([un -2*un un],[-1 0 1],NG-1,NG-1);
dphi=Poisson\(divJ(1:NG-1)*dx^2*DT);

% Linear Field Solver - Krylov solver implementation
%dphi=gmres(Poisson,divJ(1:NG-1)*dx^2*DT,10,1e-3,100);

% Impose boundary conditions
dphi=[dphi;0];

dE=zeros(NG+1,1);
dE(2:NG) = -(dphi(2:NG)-dphi(1:NG-1))/dx;
dE(1) = -(dphi(1)-dphi(NG))/dx;
dE(NG+1) = dE(1);

% residual for the average velocity
fraz=[(fraz1);1-fraz1];	
mat=sparse(p,g,fraz,N,NG);
res(1:N) = xkrylov(1:N) - v0 - 0.25*mat*QM*(2*E0(1:NG)+dE(1:NG))*DT;



\end{lstlisting}

\section*{Acknowledgments}
 This work is funded by the USAF EOARD Grant No.FA2386-14-1-5002.
  EC acknowledges support from the Leverhulme Research Project Grant Ref. 2014-112 and would like to thank Dr. Cesare Tronci for the enlighting discussions.

 

\end{document}